\documentclass[aps,pra,showkeys,nofootinbib,superscriptaddress,12pt]{revtex4-1}
\usepackage{hyperref}
\usepackage{graphicx,subfigure}
\usepackage{xcolor}
\usepackage{amsmath,amsfonts,amssymb}
\usepackage{braket}
\usepackage{dsfont}
\usepackage{todonotes}

\newcommand{\tcr}[1]{\textcolor{red}{#1}}

\begin{document}

\title{Singular spin-flip interactions for the 1D Schr\"{o}dinger operator}
%
\author{V.~Kulinskii}
\email{kulinskij@onu.edu.ua}
\affiliation{Department of Theoretical
Physics and Astronomy, Odesa National University, Dvoryanskaya 2,
65026 Odesa, Ukraine}
\author{D.~Yu. Panchenko}
\affiliation{Department of Theoretical
Physics and Astronomy, Odesa National University, Dvoryanskaya 2,
65026 Odesa, Ukraine}
\affiliation{Department of Fundamental Sciences, Odesa Military Academy, 10~Fontanska Road, Odessa 65009, Ukraine}
\email{dpanchenko@onu.edu.ua}
\begin{abstract}
We consider singular self-adjoint extensions of one-dimensional Schr\"{o}dinger operator acting in space of two-component wave functions within the framework of the distribution theory \cite{funcan_deltadistr_kurasov_jmathan1996}. We show that among $\mathds{C}^{4}$-parameter set of boundary conditions with state mixing there is only $\mathds{R}^2$-parameter subset compatible with the spin interpretation of the two-component structure of wave function. They can be identified as the point-like spin-momentum (Rashba) interactions. We suggest their physical realizations based on the regularized form of the Hamiltonian with coupling of the electrical field inhomogeneity of a background and spin of a carrier.
\end{abstract}
\maketitle
\section*{Introduction}
Point-like interactions for the Schr\"{o}dinger operators with inclusion of spin degree of freedom are of great interest for nanotechnology. Spin dependent interactions like the Rashba spin-momentum coupling \cite{qm_rashbaorigin_physolid1960} essentially influence spin dynamics and coherence in materials where such interactions are not small because of specific structure (see \cite{qm_rashbaham_nat2015} and reference therein). In low energy limit when the scale of particle delocalization exceeds all other scales including the range of interaction, one may simplify the interactions considering the point-like potential or inverse scattering length approximation \cite{book_bazeldovichperelomov_en,book_demkovostrvsk_en,bookmath_exactsolvdelta}.
General approach for the description of singular extensions of hermitian (symmetric) operators is well developed area of operator theory which was initiated by the Faddeev\&Berezin paper \cite{funcan_deltaberezinfaddeev_dan1961en} (see also \cite{bookmath_exactsolvdelta,funcan_alberverio_singperturb} and the bibliography therein). In particular, singular point-like interactions for the Schr\"{o}dinger operator
\begin{equation}\label{freeham}
    \hat{H}_{0} = -\frac{d^2}{d\,x^2}\,,
\end{equation}
form 4-parameter set $\mathds{R}^4$  of operators \cite{funcan_deltadistr_kurasov_jmathan1996,funcan_alberverio_singperturb}. They all can be derived right from the Dirac operator in one dimension \cite{funcan_deltapointdiracdabrowski_lettmp1994,qm_deltamassjumpdirac_jmph2015}. Besides well known $\delta$-potential there are extensions which can be useful for modelling the layered inhomogeneous materials \cite{qm_massjumphetero_prb1995,qm_deltamassjump_physlett2007,qm_deltasingnieto_physconferser2017}.
Physical interpretation of these interactions has been proposed in \cite{qm_deltapointsymme_physb2015,qm_deltapointmassjump_us_aoph2019}. In this respect spin-dependent interactions can appear as relativistic corrections \cite{book_ll4_en}.
But up to now spin-dependent interactions and their representation within singular perturbation of hermitian operators have not been thoroughly analyzed from the physical point of view.  In the meantime it is clear that taking into account the spin enriches the number of  point-like interactions \cite{qm_abeffectspinsing_prl1990, funcan_abeffectdabrowski_jmathphys1998}. This is equivalent to consideration of operator in extended space $L_2\oplus L_2$ and therefore it doubles the number of extensions in comparison with the 1-component case.

The generalization of known results on the point-like interactions in 1D case onto spin $s=1/2$ case using the probability current conservation seems rather straightforward  \cite{qm_deltapointrashbaus_frph2019}. But it is not necessary to interpret the two-component structure
of the wave function in term of spin only since other pseudo-spin interpretations are possible e.g. in the case of states of particle-hole elementary excitation.
In contrast to spin case the conjugate field may not have direct physical sense like magnetic field conjugated to the spin. That is why not all extensions allow interpretation as spin dependent interactions.

In this paper we derive general form of the  Hamiltonian Eq.~\eqref{freeham} with singular interactions along with corresponding boundary conditions for two-component wave function. Our technique is based on the theory of singular perturbations of finite rank for differential operators \cite{funcan_alberverio_singperturb} from which we derive the corresponding results for spin $s=1/2$ case as specific example. We also discuss the spin-flip mechanism for spin $s=1/2$ particle which has been proposed recently in \cite{qm_deltapointrashbaus_frph2019}. In Section~\ref{sec_regularizedham} we discuss the regularized form of the point-like interaction Hamiltonian. On this basis in Sectin~\ref{sec_rashba} we show that the point-like interactions of the spin-flip type can be interpreted in terms of spin-momentum (Rashba) interaction.

\section{Singular interactions for the Schr\"{o}dinger operator of spinor wave function}\label{sec_spinsinginteractions}
The boundary conditions at the singular point for the hermitian (symmetric) differential operator Eq.~\eqref{freeham} making it self-adjoint operator in $L_2$ can be found easily by imposing the condition of conservation of probability current.
In case of 2-component (spinor) wave function the Schr\"{o}dinger operator Eq.~\eqref{freeham} acts on $L_{2}\bigoplus L_{2}$
\begin{equation}\label{eq_psi2}
    \Psi = \begin{pmatrix}
    \psi_{1}\\
    \psi_{2}
    \end{pmatrix}.
\end{equation}
Then the probability current for Eq.~\eqref{freeham}  has the following matrix form:
\begin{eqnarray}\label{j_spinz}
    j=\frac{1}{i}\Gamma^{\dag}\,{\rm J_4}\,\Gamma\,,
\end{eqnarray}
where
\begin{equation}\label{bc_4vector}
  \Gamma = \begin{pmatrix}
             \psi_{1} \\
             \psi'_{1} \\
             \psi_{2} \\
             \psi'_{2} \\
           \end{pmatrix}
\end{equation}
and ${\rm J}_4$ is the block diagonal skew-symmetric matrix
\begin{eqnarray*}
    {\rm J}_4=
    \left(
        \begin{array}{cc}
            J_2& \mathbf{0}\\
            \mathbf{0} & J_2
        \end{array}
    \right),  \quad  J_2 =\left(\begin{array}{cc}
      0   & -1 \\
       1  & 0
    \end{array}\right).
\end{eqnarray*}
Therefore we introduce 4-vector (bispinor) of the boundary values at the singular point:
\begin{equation}\label{bc_4vector_pm}
  \Gamma_{0\pm 0} = \begin{pmatrix}
             \psi_{1} \\
             \psi'_{1} \\
             \psi_{2} \\
             \psi'_{2} \\
           \end{pmatrix}_{0\pm 0}
\end{equation}
and boundary condition $4\times4$-matrix $M$:
\begin{equation}\label{bc_mgamma4}
  \Gamma_{0+0} = M\,\Gamma_{0-0}\,.
\end{equation}
In terms of the matrix $M$ the conservation of probability has the form:
\begin{eqnarray}\label{Sp_2}
    {\rm J}_4=M^{\dag}{\rm J}_4\,M\,,
\end{eqnarray}
that is $M$ should be a ${\rm J}_4$ - unitary matrix. We consider the extensions (boundary conditions) which depend continuously on the interaction parameters determining the discontinuity values of function and its derivative. We can built the matrices which belong to the connected neighborhood of the unit matrix, i.e. $M = e^{t\,X}$.  From Eq.~\eqref{Sp_2} we conclude that the generators of the  corresponding Lie algebra are determined by the equation:
\begin{equation}\label{bc_generator}
   X^{\dag}\,{\rm J}_4+{\rm J}_4\,X = 0\,.
\end{equation}
Factorized solutions of Eq.~\eqref{bc_generator} correspond to non interacting, separable dynamics of wave function components.
The BC matrices intertwining the components are:
\begin{equation}\label{bc_mtrxsol}
  M = \left\{
             \begin{pmatrix}
             1&0&0&0 \\
             0&1&0&z_{1} \\
             -z^{*}_{1}&0&1&0 \\
             0&0&0&1 \\
           \end{pmatrix}\,,
\quad
           \begin{pmatrix}
             1&0&0&0 \\
             0&1&z_{3}&0 \\
             0&0&1&0 \\
             z^{*}_{3}&0&0&1 \\
           \end{pmatrix}\,,
\quad
           \begin{pmatrix}
             1&0&z_{2}&0 \\
             0&1&0&0 \\
             0&0&1&0 \\
             0&-z^{*}_{2}&0&1 \\
           \end{pmatrix}\,,
\quad
     \begin{pmatrix}
             1&0&0&z_{4} \\
             0&1&0&0 \\
             0&z^{*}_{4}&1&0 \\
             0&0&0&1 \\
           \end{pmatrix}
           \right\}\,,
\end{equation}
\begin{equation}\label{bc_psi}
\begin{pmatrix}
             \psi_{1} \\
             \psi'_{1} \\
             \psi_{2} \\
             \psi'_{2} \\
           \end{pmatrix}_{0+0}
           =
\begin{pmatrix}
  \psi_{1} \\
\psi'_{1} + z_{1}\,\psi'_{2}\\
 \psi_{2}-z^{*}_{1}\,\psi_{1}\\
 \psi'_{2} \\
\end{pmatrix}_{0-0}
,
\begin{pmatrix}
  \psi_{1}+z_{2}\,\psi_{2} \\
\psi'_{1} \\
 \psi_{2}\\
 \psi'_{2} - z^{*}_{2}\,\psi'_{1} \\
\end{pmatrix}_{0-0},
\begin{pmatrix}
  \psi_{1} \\
\psi'_{1} + z_{3}\,\psi_{2}\\
 \psi_{2}\\
 \psi'_{2} + z^{*}_{3}\,\psi_{1} \\
\end{pmatrix}_{0-0},
\begin{pmatrix}
  \psi_{1}+z_{4}\,\psi'_{2} \\
\psi'_{1} \\
 \psi_{2} +z^{*}_{4}\,\psi'_{1}\\
 \psi'_{2} \\
\end{pmatrix}_{0-0}
\end{equation}
with $z_{i}\in \mathbb{C}$ so we have $\mathbb{C}^4$ set of extensions of point-like interactions with the exchange between components of state vector $\Psi$.
\subsection{Singular distribution (Kurasov) approach}\label{subsec_kurasovsing}
The basic representation of the singular interactions of the Schr\"{o}dinger operator of a free particle in one dimension was given by P.~Kurasov in \cite{funcan_deltadistr_kurasov_jmathan1996} (see also \cite{funcan_alberverio_singperturb}). In simple terms the result is that for the operator Eq.~\eqref{freeham} all possible BC's at the singular point  describe the jumps in wave function and  its derivative. So one comes to the consideration of the functionals on the space of bounded functions and their derivatives outside the singular point ($x=0$) with possible discontinuity at the singular point. Here we adjust the general theory of \cite{funcan_alberverio_singperturb} to the case of the Schr\"{o}dinger operator which
acts in the Hilbert space of 2-component wave Eq.~\eqref{eq_psi2}.

Following the general scheme of \cite{funcan_alberverio_singperturb} we define the following distributions for the components:
\begin{align}\label{eq_diracdeltaspin}
    \delta_{i}\left[ \Psi \right] =&
    \frac{
    \psi_{i}(0-)+\psi_{i}(0+)}{2}\,,\quad
    \delta^{(1)}_{i}\left[ \Psi \right] = -
    \frac{
    \psi'_{i}(0-)+\psi'_{i}(0+)}{2}\,,\\
    \beta_{i}\left[ \Psi \right] =&
    \frac{
    \psi_{i}(0+)-\psi_{i}(0-)}{2}\,,\quad
    \beta^{(1)}_{i}\left[ \Psi \right] = -
    \frac{
    \psi'_{i}(0+)-\psi'_{i}(0-)}{2}
\end{align}
and introduce the vector-functionals
\[\Delta_{i} = \begin{pmatrix}
\delta_{i}\\
\delta^{(1)}_{i}\\
\end{pmatrix}.\]
Then the operator expression
\begin{equation}\label{eq_kurasovsinghamspin}
     \hat{H}_{s} =
\begin{pmatrix}
        -\,D^{2}_{x}&0\\
    0&-\,D^{2}_{x}
    \end{pmatrix}
+
    \begin{pmatrix}
    \Delta^{T}_{1}(.)\,h^{1}_{1}\,\Delta_{1} &\, \Delta^{T}_{2}(.)\,h^{1}_{2}\,\Delta_{1}\\
    \Delta^{T}_{1}(.)\,h^{2}_{1}\,\Delta_{2}&\,
    \Delta^{T}_{2}(.)\,h^{2}_{2}\,\Delta_{2}
    \end{pmatrix}
\end{equation}
determines the Hamiltonian of a free particle with augmented with the point-like interactions. Eq.~\eqref{eq_kurasovsinghamspin} defines finite rank perturbation of the Schr\"{o}dinger operator (see formal definitions of corresponding functional spaces in \cite{funcan_alberverio_singperturb}).
The second term in Eq.~\eqref{eq_kurasovsinghamspin}) is the hermitian quadratic form of linear functionals.
The matrices
$h_{i,j}$ of size $2 \times 2$ form the hermitian block matrix:
\begin{equation}\label{h_mtrx}
    \mathbf{h} = \begin{pmatrix}
    h^{1}_{1} &h^{1}_{2}\\
    h^{2}_{1}&h^{2}_{2}
    \end{pmatrix}.
\end{equation}
The relation between matrices $\mathbf{h}$ and the boundary value matrices $M$ can be found using the general theory \cite{funcan_alberverio_singperturb} and is as following:
\begin{equation}\label{eq_hm}
M = \left(\mathds{1} -
    A\,\mathbf{h}\,I\right)^{-1}\,(\mathds{1} +
    A\,\mathbf{h}\,I) 
\end{equation}
where in specific case of two-component wave function Eq.~\eqref{eq_psi2}:
\[A = \frac{1}{2}\,\begin{pmatrix}
\,0\,&\,1\,&\,0\,&\,0\,\\
1&0&0&0\\
0&0&0&1\\
0&0&1&0
\end{pmatrix}\,,\quad
I =\begin{pmatrix}
1&0&0&0\\
0&-1&0&0\\
0&0&1&0\\
0&0&0&-1
\end{pmatrix}.
\]
%
So using Eq.~\eqref{eq_hm} we find that the BC matrices Eq.~\eqref{bc_mtrxsol} with state mixing  correspond to the following singular interaction matrices in Eq.~\eqref{eq_kurasovsinghamspin}:
\begin{equation}\label{bc_htrxsol}
  \mathbf{h} = \left\{
  \begin{pmatrix}
             0&0&0&-z_{1} \\
             0&0&0&0 \\
             0&0&0&0 \\
             -z^*_{1}&0&0&0 \\
           \end{pmatrix}\,,
\quad
\begin{pmatrix}
             0&0&0&0 \\
             0&0&0&-z_{2} \\
             0&0&0&0 \\
             0&-z^*_{2}&0&0 \\
           \end{pmatrix}\,, \quad
           \begin{pmatrix}
             0&0&z_{3}&0 \\
             0&0&0&0 \\
             z^*_{3}&0&0&0 \\
             0&0&0&0 \\
           \end{pmatrix}\,,
\quad
           \begin{pmatrix}
             0&0&0&0 \\
             0&0&z_{4}&0 \\
             0&z^*_{4}&0&0 \\
             0&0&0&0 \\
           \end{pmatrix}
           \right\}\,.
\end{equation}
In the following Section we build the regularized form of the singular Hamiltonian Eq.~\eqref{eq_kurasovsinghamspin}. Such a form is useful for physical realization of point-like interactions which are nothing but limiting cases of spatial inhomogeneities of specific physical fields \cite{qm_deltapointrashbaus_frph2019}.

\section{Regularized form of point-like interactions}\label{sec_regularizedham}
Basing on the results of previous Section and applying technique of \cite{funcan_deltadistr_kurasov_jmathan1996} we can give explicit regularized form of the Hamiltonian. Let us define the linear space $K$ of test functions which are continuous and bounded along with their derivatives outside 0 (see details of formal definitions in \cite{funcan_deltadistr_kurasov_jmathan1996,funcan_alberverio_singperturb}). Let $K'$ is the dual linear space i.e. the linear space of distributions on $K$.
In two-component case we consider the differential operator
\begin{equation}
    \hat{H}_{s} = \begin{pmatrix}
        H_{1,1}& H_{1,2}\\
        H_{2,1}& H_{2,2}
    \end{pmatrix}
\end{equation}
which can be represented as the following:
\begin{multline}\label{ham_d2kurasov}
    \hat{H}_{s} =
    \left(i\,D_{x}\right)^{2}\begin{pmatrix}
    1+\mathbf{a}_{1,1}\,\Delta_{1,1}&\mathbf{a}_{1,2}\,\Delta_{1,2}\\
    \mathbf{a}_{2,1}\,\Delta_{2,1}&1+\mathbf{a}_{2,2}\,\Delta_{2,2}
    \end{pmatrix}
    +
    \left(i\,D_{x}\right)
    \begin{pmatrix}
    \mathbf{b}_{1,1}\,\Delta_{1,1}&\mathbf{b}_{1,2}\,\Delta_{1,2}\\
    \mathbf{b}_{2,1}\,\Delta_{2,1}&\mathbf{b}_{2,2}\,\Delta_{2,2}
    \end{pmatrix}
    +\\
    \begin{pmatrix}
    \mathbf{c}_{1,1}\,\Delta_{1,1}&\mathbf{c}_{1,2}\,\Delta_{1,2}\\
    \mathbf{c}_{2,1}\,\Delta_{2,1}&\mathbf{c}_{2,2}\,\Delta_{2,2}
    \end{pmatrix}\,,
\end{multline}
where vectors $\mathbf{a},\mathbf{b},\mathbf{c}$ determine the parameters of corresponding singular interactions.
This operator acts on the corresponding Sobolev space $W^{2}_{2}(\mathbb{R}\setminus 0)\oplus W^{2}_{2}(\mathbb{R}\setminus 0)$ and maps it into linear space of functionals $K' \oplus K'$ due to actions of the following operators
\begin{equation}\label{eq_deltaijdef}
    \Delta_{i,j} = \begin{pmatrix}
                    \delta_{i,j}\\
                    \delta^{(1)}_{i,j}
    \end{pmatrix}\,.
\end{equation}
Then the product of $\delta$-distribution and its derivative (the elements of $K'$) and the elements $\psi \in K$ can be written as:
\begin{align}
\delta_{i,j}\,\psi = &\,\, \delta_{i}[\psi]\,\delta_{j} + \beta_{i}[\psi]\,\beta_{j}\\
\delta^{(1)}_{i,j}\,\psi = &\,\, \delta^{(1)}_{i}[\psi]\,\delta_{j} + \beta^{(1)}_{i}[\psi]\,\beta_{j}+\delta_{i}[\psi]\,\delta_{j}^{(1)}+\beta_{i}[\psi]\,\beta_{j}^{(1)}
\end{align}
which are analogues to the expressions of scalar case (see \cite{funcan_deltadistr_kurasov_jmathan1996,funcan_alberverio_singperturb}).
Being restricted to the subspace of continuous functions  Eq.~\eqref{ham_d2kurasov} maps it onto 4-dim subspace of distributions spanned by the functionals $\delta^{(k)},\, k = 0,1,2,3$.
Imposing the constraint $L\Psi \in L_{2}\oplus L_{2}$ we get the relations
\begin{equation}\label{eq_d2linsyskurasov}
\begin{pmatrix}
  -a^{(1)}_{1,1}\, & 0 & 0 & 0 & -a^{(1)}_{1,2} & 0 & 0 & 0 \\
 i b^{(1)}_{1,1}-a^{(0)}_{1,1}\, & -a^{(1)}_{1,1} & 0 & 0 & i b^{(1)}_{1,2}-a^{(0)}_{1,2} & -a^{(1)}_{1,2} & 0 & 0 \\
 i b^{(0)}_{1,1}+c^{(1)}_{1,1}\, & i b^{(1)}_{1,1} & -2 & 0 & i b^{(0)}_{1,2}+c^{(1)}_{1,2} & i b^{(1)}_{1,2} & 0 & 0 \\
 c^{(0)}_{1,1} & c^{(1)}_{1,1}\, & 0 & 2 & c^{(0)}_{1,2} & c^{(1)}_{1,2} & 0 & 0 \\
 -a^{(1)}_{2,1}\, & 0 & 0 & 0 & -a^{(1)}_{2,2} & 0 & 0 & 0 \\
 i b^{(1)}_{2,1}-a^{(0)}_{2,1} & -a^{(1)}_{2,1} & 0 & 0 & i b^{(1)}_{2,2}-a^{(0)}_{2,2} & -a^{(1)}_{2,2} & 0 & 0 \\
 i b^{(0)}_{2,1}+c^{(1)}_{2,1}\, & i b^{(1)}_{2,1} & 0 & 0 & i b^{(0)}_{2,2}+c^{(1)}_{2,2} & i b^{(1)}_{2,2} & -2 & 0 \\
 c^{(0)}_{2,1}\, & c^{(1)}_{2,1} & 0 & 0 & c^{(0)}_{2,2} & c^{(1)}_{2,2} & 0 & 2
\end{pmatrix}\,
\begin{pmatrix}
\delta_{1}[\psi]\\
\delta^{(1)}_{1}[\psi]\\
\beta_{1}[\psi]\\
\beta^{(1)}_{1}[\psi]\\
\delta_{2}[\psi]\\
\delta^{(1)}_{2}[\psi]\\
\beta_{2}[\psi]\\
\beta^{(1)}_{2}[\psi]
\end{pmatrix}=0\,.
\end{equation}
The boundary conditions follow from the demand that the rank of the matrix in Eq.~\eqref{eq_d2linsyskurasov} is equal to 4. Therefore we get:
\begin{align}
    a^{(1)}_{1,1} = & \, 0\,,\,\, i\,b^{(1)}_{1,1} = a^{(0)}_{1,1}\,,
    a^{(1)}_{1,2}=  \, 0\,,\,\, i\,b^{(1)}_{1,2} = a^{(0)}_{1,2}\\
    a^{(1)}_{2,1} = & \, 0\,,\,\, i\,b^{(1)}_{2,1} = a^{(0)}_{2,1}\,,
    a^{(1)}_{2,2}=  \, 0\,,\,\, i\,b^{(1)}_{2,2} = a^{(0)}_{2,2}
\end{align}
and the corresponding BC's are:
\begin{equation}\label{eq_betadelta4}
2\,\begin{pmatrix}
\beta_{1}[\psi]\\
\beta^{(1)}_{1}[\psi]\\
\beta_{2}[\psi]\\
\beta^{(1)}_{2}[\psi]
\end{pmatrix} =
\Lambda\,
\begin{pmatrix}
\delta_{1}[\psi]\\
\delta^{(1)}_{1}[\psi]\\
\delta_{2}[\psi]\\
\delta^{(1)}_{2}[\psi]
\end{pmatrix}\,,
\end{equation}
where
\begin{equation}
    \Lambda = \begin{pmatrix}
 i b^{(0)}_{1,1}+c^{(1)}_{1,1} & i b^{(1)}_{1,1} & i b^{(0)}_{1,2}+
 c^{(1)}_{1,2}&
   i b^{(1)}_{1,2} \\
 -c^{(0)}_{1,1} & -c^{(1)}_{1,1} & -c^{(0)}_{1,2} & -c^{(1)}_{1,2}\\
 i b^{(0)}_{2,1}+c^{(1)}_{2,1} & i b^{(1)}_{2,1} & i b^{(0)}_{2,2}+
 c^{(1)}_{2,2} &
 i b^{(1)}_{2,2}\\
 -c^{(0)}_{2,1}& -c^{(1)}_{2,1}& -c^{(0)}_{2,2}& -c^{(1)}_{2,2}\\
\end{pmatrix}.
\end{equation}
As long as we are interested in spin-flip BC's only we
may put diagonal coefficients to zero so that Eq.~\eqref{eq_betadelta4} becomes as following:
\begin{equation}\label{eq_betadelta40}
2\,\begin{pmatrix}
\beta_{1}[\psi]\\
\beta^{(1)}_{1}[\psi]\\
\beta_{2}[\psi]\\
\beta^{(1)}_{2}[\psi]
\end{pmatrix} =
\begin{pmatrix}
 0 & 0 & i b^{(0)}_{1,2}+
 c^{(1)}_{1,2}&i b^{(1)}_{1,2} \\
 0 & 0 & -c^{(0)}_{1,2} & -c^{(1)}_{1,2}\\
 i b^{(0)}_{2,1}+c^{(1)}_{2,1} & i b^{(1)}_{2,1} &0 &
 0\\
 -c^{(0)}_{2,1}& -c^{(1)}_{2,1}& 0& 0
\end{pmatrix}
\begin{pmatrix}
\delta_{1}[\psi]\\
\delta^{(1)}_{1}[\psi]\\
\delta_{2}[\psi]\\
\delta^{(1)}_{2}[\psi]
\end{pmatrix}.
\end{equation}

The current conservation Eq.~\eqref{j_spinz} yields additional constraints
\begin{align}\label{eq_coeff4bc}
    c^{(0)}_{2,1} =&\,   \hphantom{ -}c^{(0)^{*}}_{1,2}\,,\quad
    &&c^{(1)^{*}}_{2,1} = c^{(1)}_{1,2}+i\,b^{(0)}_{1,2}\,, \notag\\
    b^{(1)}_{2,1} =& - b^{(1)^{*}}_{1,2}\,,\quad  &&b^{(0)}_{2,1}\hphantom{^{*}} =b^{(0)^{*}}_{1,2}.
\end{align}
From Eq.~\eqref{eq_betadelta40}, \eqref{eq_coeff4bc} we finally
obtain $\mathds{C}^4$ set of spin-flip BC's matrices:
\begin{equation}\label{bc_lambdasol}
  \Lambda = \left\{
            \begin{pmatrix}
             0&0&0&0 \\
             0&0&-z_{1}&0 \\
             0&0&0&0 \\
             -z^{*}_{1}&0&0&0 \\
           \end{pmatrix}\,, \quad
           \begin{pmatrix}
             0&0&z_2&0 \\
             0&0&0&0 \\
             0&0&0&0 \\
             0&-z^{*}_{2}&0&0 \\
           \end{pmatrix}\,,
\quad
           \begin{pmatrix}
             0&0&0&0 \\
             0&0&0&-z_{3} \\
             z^{*}_{3}&0&0&0 \\
             0&0&0&0 \\
           \end{pmatrix}\,,
           \quad
\begin{pmatrix}
             0&0&0&z_{4} \\
             0&0&0&0 \\
             0&z^{*}_{4}&0&0 \\
             0&0&0&0 \\
           \end{pmatrix}
           \right\}\,.
\end{equation}
With account of the relation
\[M =\left(I+\Lambda\right)\,\left(I-\Lambda\right)^{-1} \]
Eq.~\eqref{bc_lambdasol} leads exactly to Eq.~\eqref{bc_mtrxsol}.
Thus the Hamiltonian has the form
\begin{multline}\label{ham_kurasovspinflip}
    \hat{H}_{s} =
    \left(i\,D_{x}\right)^{2}\begin{pmatrix}
    1&i\,z_{4}\,\delta\\
    -i\,z^{*}_{4}\,\delta&1
    \end{pmatrix}
    +
    \left(i\,D_{x}\right)
    \begin{pmatrix}
    0&-i\left(z_2+z_3\right)\,\delta+z_{4}\,\delta^{(1)}\\
    i\left(z^{*}_2+z^{*}_3\right)\,\delta-z^{*}_{4}\,\delta^{(1)}&0
    \end{pmatrix}
    +\\
    \begin{pmatrix}
    0&z_{1}\,\delta + \tcr{z_{3}}\,\delta^{(1)}\\
    z^*_{1}\,\delta+z^{*}_{2}\,\delta^{(1)}&0
    \end{pmatrix}.
\end{multline}
Other solutions represent the BC's without state mixing and correspond to known cases \cite{funcan_deltadistr_kurasov_jmathan1996,bookmath_exactsolvdelta} for each of the components. We will not consider them thoroughly and just give one example of relevant physical situation which could be of interest in this case. 

Having established the BC's in 1 dimension we can apply these results to 3D case. The latter can be effectively reduced to one dimensional problem on semi-axis $r>0$ with  $\phi(r) = r\,\psi(r)$ as the effective 1D wave function through the natural definition domain of free Hamiltonian:
\begin{equation}\label{eq_3dham0}
    \vert\vert\hat{H}_{0}\Psi\vert\vert^2 = \int\limits_{0}^{\infty} \vert(r\,\psi)''\vert^2 d\,r<\infty\,.
\end{equation}
Then the limiting value $\phi(0)$ as well as its derivative $\phi'(0)$ is defined since Eq.~\eqref{eq_3dham0} is well defined  on the corresponding Sobolev space $W^{2}_{2}(\mathbb{R}_{+})$ which is dense in $L_2(\mathbb{R}_{+})$.
The conservation of the probability current:
\begin{equation}\label{j_3d}
    J = \int\limits_{0}^{\infty} {\rm Im} \left(\Psi^{\dagger}\,\frac{\partial \Psi}{\partial r}\right) r^2\,dr =
    \int\limits_{0}^{\infty} {\rm Im} \left((r\Psi^{\dagger})\,\frac{\partial (r \Psi)}{\partial r}\right)\,dr
\end{equation}
leads to the BC's of the following form:
\begin{equation}\label{eq_wbc}
\Phi' = W\,\Phi\,,
\end{equation}
where $W$ is the hermitean matrix and:
\begin{equation}\label{bc_3dspin}
    \Phi = \left(
        \begin{array}{c}
            \phi_{\uparrow}\\
            \phi_{\downarrow}
        \end{array}
    \right)\,,\quad
    \Phi' =
    \left(
        \begin{array}{c}
            \phi'_{\uparrow}\\
            \phi'_{\downarrow}
        \end{array}
    \right)
\end{equation}
are 2-spinor boundary-value vectors.
The standard representation:
\begin{equation}\label{paulidecomposition}
    W = \Omega\,I + \mathbf{w}\cdot \boldsymbol{\sigma}\,,
\end{equation}
where $\sigma$ is the Pauli matrices vector allows to identify
the scalar part (first term) as the standard point-like potential b.c. \cite{funcan_deltaberezinfaddeev_dan1961en,book_bazeldovichperelomov_en}:
\begin{equation}\label{bc_standardelta}
     \Phi' =  \Omega\,\Phi
\end{equation}
independent on the spin state.
The vector part (traceless second term) of Eq.~\eqref{paulidecomposition} describes polarizational contact interactions. Namely spin-dependent version of Eq.~\eqref{bc_standardelta} is:
\begin{equation}\label{bc_deltaspin}
    \Phi' =  \omega\,
    \left(
        \begin{array}{cc}
            1&0\\
            0&-1
        \end{array}
    \right)\,\Phi\,,
\end{equation}
where the bound state exists only for one component (e.g. for $\downarrow$-state if $\omega>0$ ). Combining this case with the standard ZRP Eq.~\eqref{bc_standardelta} with $\Omega<0,\,\,|\Omega|>|\omega|$)
it is possible to get two bound states with
\begin{equation}\label{enrg_hyperfinesplit}
E_{\uparrow,\downarrow} \sim -(\Omega\pm \omega)^2
\end{equation}
for $\uparrow,\downarrow$ - states. Thus Eq.~\eqref{enrg_hyperfinesplit}
can be treated as the hyperfine splitting due to Fermi contact interaction \cite{qm_fermispinnucl_zph1930} between the magnetic moments of an ``electron`` and a ``nucleus``, i.e. the singular center.


\section{Spin-momentum (Rashba) point-like interactions}\label{sec_rashba}
As has been mentioned above two-component structure of the wave function does not necessary mean the consideration of spin particle. Spin interpretation implies additional constraint on the structure of probability current.
In fact in such physical situation we deal with the Pauli Hamiltonian:
\begin{equation}\label{eq_ham_pauli}
\hat{H} = \frac{\left(\hat{p} - \frac{q}{c}\,\mathbf{A}\right)^2}{2\,m} + q\,\varphi - \frac{q\,\hbar}{2\,m\,c}\,\hat{\boldsymbol{\sigma}}\cdot\vec{\mathcal{H}}
\end{equation}
with $\boldsymbol{\sigma}$ representing the vector of Pauli matrices
and $\vec{\mathcal{H}}$ is the magnetic field with $\mathbf{A}$
as its vector potential and $\varphi$ being the scalar potential.

The probability current for \eqref{eq_ham_pauli} is as following:
\begin{eqnarray}\label{j_spinpauli}
{\rm \mathbf{J}}_{w} = {\frac{\hbar} {m}}{\rm
Im}\left( {\Psi^{\dagger} \nabla \Psi} \right) - \frac{q}{mc}{\rm
\mathbf{A}}\Psi^{\dagger} \Psi + {\frac{\hbar}{2m}}{\rm rot}\left({\Psi^{\dagger} \boldsymbol{\sigma} \,\,\Psi}  \right)\,,
\end{eqnarray}
where the last term is due to the magnetization current (see e.g. \cite{book_ll4_en}).
The constraint of conservation of Eq.~\eqref{j_spinpauli} reduces the number of BC's
with the spin-flip to only two-parameter set $\mathds{R}^2$  because of the conservation constraints for $y,z$ components of \eqref{j_spinpauli}  (see details in
\cite{qm_deltapointrashbaus_frph2019}):
\begin{eqnarray}\label{jxyz_spin}
J_{y}=&-\left(\frac{\partial \Psi^\dagger}{\partial x} \sigma_{z}\Psi +\Psi^\dagger\sigma_{z}\frac{\partial \Psi}{\partial x}\right)\,,\\
J_{z}=&\frac{\partial \Psi^\dagger}{\partial x} \sigma_{y}\Psi +\Psi^\dagger\sigma_{y}\frac{\partial \Psi}{\partial x}\,.
\end{eqnarray}
Here we demonstrate the same result directly
basing on the structure of the Hamiltonian Eq.~\eqref{ham_kurasovspinflip}. Let us consider the simplest specific example  of $\delta$-interaction with the spin-flip ($z_1$-coupling)
in regularized form using the spin representation for Eq.~\eqref{ham_kurasovspinflip}:
\begin{align}\label{ham_kurasovspinrepres}
    \hat{H}_{z_{1}} = &\left(i\,D_{x}\right)^2 + \left(a\,\sigma_{x}+b\,\sigma_{y}\right)\,V_{\varepsilon}(x)\,, \\
    a\sim & \,{\rm Re}\,z_{1}\,,\quad
    b\sim {\rm Im}\,z_{1} \notag
\end{align}
where $V_{\varepsilon}(x)$ is even smooth ($C^{\infty}$- class) function with finite support (which includes $x=0$). It represents the weakly convergent $\delta$-sequence (see details in \cite{funcan_deltadistr_kurasov_jmathan1996}). As long as the spin-flip operator should be of the form $\vec{\mathcal{H}}\cdot \hat{\mathbf{S}}$ with ${\rm div}\,\vec{\mathcal{H}} = 0$ we see that the only physical choice in Eq.~\eqref{ham_kurasovspinrepres} is $a=0$. The same reasoning works for $z_{4}$-coupling. Therefore we get the Hamiltonian with the singular spin-flip interactions:
\begin{equation}\label{ham_rx1x4}
    \hat{H}_{z_{4}} = \left(i\,D_{x}\right)^2
    \left(I + X^{(r)}_{4}\sigma_{y}\,\delta\right)+ i\,D_{x}\left(-i\, X^{(r)}_{4}\sigma_{y}\,\delta^{(1)}\right)+
    X^{(r)}_{1}\sigma_{y}\,\delta
\end{equation}
with
\[X^{(r)}_{1} = -\, {\rm Im}\, z_{1}\,,\quad
X^{(r)}_{4} = -\, {\rm Re}\, z_{4}\,.
\]
Substituting these results in Eq.~\eqref{bc_psi} we get the following BC's:
\begin{equation}\label{mr_matrix1}
\begin{pmatrix}
             \psi_{\uparrow} \\
             \psi'_{\uparrow} \\
             \psi_{\downarrow} \\
             \psi'_{\downarrow} \\
           \end{pmatrix}_{0+0}
           =
\begin{pmatrix}
  \psi_{\uparrow} \\
\psi'_{\uparrow} + 2\,X^{(r)}_{1}\,\psi'_{\downarrow}\\
 \psi_{\downarrow}\\
 \psi'_{\downarrow}+ 2\,X^{(r)}_{1}\,\psi'_{\uparrow}\\
\end{pmatrix}_{0-0}
\end{equation}
and
\begin{equation}\label{mr_matrix4}
\begin{pmatrix}
             \psi_{\uparrow} \\
             \psi'_{\uparrow} \\
             \psi_{\downarrow} \\
             \psi'_{\downarrow} \\
           \end{pmatrix}_{0+0}
           =    \begin{pmatrix}
           \psi_{\uparrow} +
          2\,X^{(r)}_{4} \,\psi'_{\downarrow}\\
           \psi'_{\uparrow}\\
           \psi_{\downarrow}+2\,X^{(r)}_{4}\,
              \psi'_{\uparrow}  \\
              \psi'_{\downarrow}\\
              \end{pmatrix}_{0-0}.
\end{equation}
They coincide with those obtained in \cite{qm_deltapointrashbaus_frph2019}. Thus the regularized form of the spin-flip Hamiltonian is:
\begin{equation}\label{ham_regx1x4}
    \hat{H} = \left(i\,D_{x}\right)^2
    \left(I + X^{(r)}_{4}\sigma_{y}\,V_{\varepsilon}(x)\right)+ i\,D_{x}\left(-i\, X^{(r)}_{4}\sigma_{y}\,V^{(1)}_{\varepsilon}(x) \right)+
    X^{(r)}_{1}\sigma_{y}\,V_{\varepsilon}(x)
\end{equation}
which means that the spin projections $S_{i}=\frac{\hbar}{2}\,\sigma_{i}\,, i=x,z$ are not integrals of motions anymore.
The structure of the spin interaction terms in Eqs.~\eqref{ham_rx1x4} and \eqref{ham_regx1x4} allows to identify their physical origin with the relativistic
spin-momentum (Rashba) coupling (see \cite{book_ll4_en, qm_rashbabychkov_zhetp1984, qm_rashbaham_nat2015}):
\begin{equation}
    \hat{H}_{{\rm int}} = \lambda\,\left(\mathbf{p}\times \nabla \, \varphi \right)\cdot \hat{\mathbf{S}}
\end{equation}
in layered systems. Indeed, due to translational symmetry $\varphi$ depends on $x$ only and represents spatial inhomogeneity of the electrostatic potential. Thus the interpretation of two-component nature of $\Psi$ as spin degree of freedom singles 2-parameter set $\mathbb{R}^2$ out of the set $\mathbb{C}^4$ of possible extensions  Eq.~\eqref{bc_htrxsol}.
As is clear from the structure of Eq.~\eqref{ham_regx1x4} we may attribute $X^{(r)}_{1}$-coupling to the Rashba interaction of the ``bare`` carrier due to the inhomogeneous electrostatic field and transverse component of the momentum $p_z$. The same is true for $X^{(r)}_{4}$-coupling though in this case such field modifies the kinetic energy term. We may conclude that this is due to spatial dependent effective mass. Thus the BC corresponding to this singular extension can be realized in systems where the propagation of a particle in a medium lead to its ``dressing`` and dynamic contribution to the mass operator. The latter
thus expected in the materials with spatially variable effective mass of the carriers.

The point-like interaction with the spin-flip:
\begin{equation}\label{bc_spinor_deltaspinflip}
 \Phi'  = \left(
        \begin{array}{cc}
            0&z\\
            z^*&0
        \end{array}
    \right)\, \Phi\,,\quad z \in \mathbb{C}
\end{equation}
is reduced to the above considered cases Eqs.~\eqref{bc_standardelta},~\eqref{bc_deltaspin} in the basis of eigenstates of $\hat{S}_{x}$:
\[\ket{S_{x} = \pm 1/2} = \frac{\ket{\uparrow}\pm\ket{\downarrow}}{\sqrt{2}}\,.\]
\section*{Conclusion}
In this paper we have reproduced the results of \cite{qm_deltapointrashbaus_frph2019} on point-like interactions for the Schr\"{o}dinger operator Eq.~\eqref{freeham} on space of two-component wave functions $L_2\oplus L_2$. We use the Kurasov's technique for construction of singular perturbations of self-adjoint operators \cite{funcan_alberverio_singperturb} and show that there is $\mathds{C}^{4}$ set of singular interactions with interstate mixing Eq.~\eqref{bc_htrxsol}. If one uses spin interpretation of two-component wave vector then this set reduces to $\mathds{R}^{2}$ set i.e. only two spin-momentum couplings Eqs.~\eqref{mr_matrix1},\eqref{mr_matrix4}.
The coupling Eq.~\eqref{mr_matrix1} is due to localized inhomogeneity of the potential barrier. The b.c. Eq.~\eqref{mr_matrix4} is generated by the effective mass spatial dependence and is of non-local type \cite{qm_deltapointmassjump_us_aoph2019}. Besides spin-dependent point-like interactions can be realized as the thin magnetized interface in layered system.
In this respect it is important to analyze extensions for Aharonov-Bohm Hamiltonian in 2D \cite{qm_abeffectspinsing_prl1990}. Here one can expect interesting interplay between localized magnetic flux (see \cite{funcan_abeffectdabrowski_jmathphys1998}) and relativistic spin-momentum interactions  \cite{qm_spinorbitpavlov_arxive2004,funcan_spinorbit_jmp2006} which is of interest for studying low dimensional magnetic systems.
It seems that the spin-flip self-adjoint extensions which represent point-like spin-momentum coupling can be also obtained by the standard von Neumann method for the Dirac Hamiltonian similar to that used in \cite{funcan_deltapointdiracdabrowski_lettmp1994,qm_deltapointdrac_arxive2019} though the spin was not taken into account there.
The nature of other one dimensional b.c.'s deals with the pseudospin interpretations of two-component nature of the wave function, e.g. particle-hole excitation states near Fermi level. The question about physical realization of such extensions will be the subject of future work.

\section*{Acknowledgments}
This work was completed due to MES Ukraine grants 10115U003208, 10118U000202. V.K. appreciate support via Fulbright Research Grant (IIE ID: PS00245791) and is also grateful to Mr.~Konstantin Yun for long-term financing of the individual research.  Authors thank Prof. Vadim Adamyan for discussion and comments.
%
\end{document}